\documentclass[journal=jacsat,manuscript=article]{achemso}
\usepackage{xcolor}
\usepackage[version=3]{mhchem} 

\author{Hui-Fei Zhai}
\affiliation{Department of Physics, Zhejiang University, Hangzhou 310027, China}

\author{Pan Zhang}
\affiliation{Department of Physics, Zhejiang University, Hangzhou 310027, China}

\author{Si-Qi Wu}
\affiliation{Department of Physics, Zhejiang University, Hangzhou 310027, China}

\author{Chao-Yang He}
\affiliation{Department of Physics, Zhejiang University, Hangzhou 310027, China}

\author{Zhang-Tu Tang}
\affiliation{Department of Physics, Zhejiang University, Hangzhou 310027, China}

\author{Hao Jiang}
\affiliation{Department of Physics, Zhejiang University, Hangzhou 310027, China}

\author{Yun-Lei Sun}
\affiliation{Department of Physics, Zhejiang University, Hangzhou 310027, China}

\author{Jin-Ke Bao}
\affiliation{Department of Physics, Zhejiang University, Hangzhou 310027, China}

\author{I. Nowik}
\affiliation{Racah Institute of Physics, The Hebrew University,
Jerusalem 91904, Israel}

\author{I. Felner}
\affiliation{Racah Institute of Physics, The Hebrew University,
Jerusalem 91904, Israel}

\author{Yue-Wu Zeng}
\affiliation{Center of Electron Microscope, Zhejiang University, Hangzhou 310027, China}

\author{Yu-Ke Li}
\affiliation{Department of Physics, Hangzhou Normal University,
Hangzhou 310036, China}

\author{Xiao-Feng Xu}
\affiliation{Department of Physics, Hangzhou Normal University,
Hangzhou 310036, China}

\author{Qian Tao}
\affiliation{Department of Physics, Zhejiang University, Hangzhou 310027, China}

\author{Zhu-An Xu}
\affiliation{Department of Physics, Zhejiang University, Hangzhou 310027, China}
\alsoaffiliation{State Key Lab of Silicon Materials, Zhejiang University, Hangzhou 310027, China}

\author{Guang-Han Cao}
\affiliation{Department of Physics, Zhejiang University, Hangzhou 310027, China}
\alsoaffiliation{State Key Lab of Silicon Materials, Zhejiang University, Hangzhou 310027, China}
\email{ghcao@zju.edu.cn}

\title[A New Self-doped Superconductor Eu$_3$Bi$_2$S$_{4}$F$_4$]
  {Anomalous Eu Valence State and Superconductivity in Undoped Eu$_3$Bi$_2$S$_{4}$F$_4$}

\abbreviations{XRD, TEM}

\begin{document}


\begin{abstract}
  We have synthesized a novel europium bismuth sulfofluoride, Eu$_3$Bi$_2$S$_4$F$_4$, by solid-state reactions in sealed evacuated quartz ampoules. The compound crystallizes in a tetragonal lattice (space group $I4/mmm$, $a$ = 4.0771(1) {\AA}, $c$ = 32.4330(6) {\AA}, and $Z$ = 2), in which CaF$_2$-type Eu$_3$F$_4$ layers and NaCl-like BiS$_2$ bilayers stack alternately along the crystallographic $c$ axis. There are two crystallographically distinct Eu sites, Eu(1) and Eu(2) at the Wyckoff positions 4$e$ and 2$a$, respectively. Our bond-valence-sum calculation, based on the refined structural data, indicates that Eu(1) is essentially divalent, whilst Eu(2) has an average valence of $\sim$ +2.64(5). This anomalous Eu valence state is further confirmed and supported, respectively, by M\"{o}ssbauer and magnetization measurements. The Eu$^{3+}$ components donate electrons into the conduction bands that are mainly composed of Bi- 6$p_x$ and 6$p_y$ states. Consequently, the material itself shows metallic conduction, and superconducts at 1.5 K without extrinsic chemical doping.
\end{abstract}

\section{Introduction}

Exploration of new superconductors is an important and challenging work. Through long-term effort, especially during the past three decades, one has learned two basic rational routes for new superconducting materials. The first one is to carry out chemical doping in a so-called parent compound which is potentially superconducting. The other is to synthesize a new compound that contains two-dimensional (2D) superconductively active layers via structural design.\cite{ferey,cario,jh} Prominent examples are manifested by copper-based and iron-based high-temperature superconductors, in which the superconductively active layers are 2D CuO$_2$ planes and Fe$_2$As$_2$ layers, respectively. Impressively, both the seminal discoveries were made by chemical doping,\cite{bednorz&muller,hosono} and all structures of these high-temperature superconductors can be well elucidated in terms of their crystal chemistry.\cite{raveau,jh} Optimizations of the doping level and the building block layers may create a record of the superconducting transition temperature, $T_\text{c}$.\cite{schilling,GdTh} It is noted here, nevertheless, superconductivity may also appears, though infrequently or even rarely, in an apparently \emph{undoped} material in which charge carriers are induced by self doping through internal charge transfer.\cite{cao,sun}

Recently, superconductivity up to 10.6 K was discovered in Bi$_4$O$_4$S$_3$\cite{mizuguchi-1} and LaO$_{1-x}$F$_{x}$BiS$_2$,\cite{mizuguchi-2} in which the 2D BiS$_2$ bilayers were identified as superconductively active layers. Thus, the BiS$_2$-based material represents a new class of superconductors, though the superconducting phase in the Bi-O-S system is not well established due to the complicated stacking faults.\cite{jpsj-BiOS,jacs-BiOS1,jacs-BiOS2} The LaBiS$_2$O$_{1-x}$F$_{x}$ (hereafter abbreviated as La-1121)\bibnote{The chemical formula should preferably written as LaBiS$_2$O$_{1-x}$F$_{x}$ according to standard nomenclature of inorganic compounds. See: Hiroi, Z. A. arXiv: \textbf{2008}, 0805.4668.} system has been extended into a superconducting family whose new members appeared one after another via element replacements in the La$_2$O$_2$ spacer layers\cite{whh,Nd1121,Pr1121,yazici-1,lyk,zhai} or even in the BiS$_2$ bilayers.\cite{BiSe2}  As a matter of fact, the relevant work can be traced back earlier to 1970s when the crystal structure of CeBiS$_2$O was first reported.\cite{Ce1121} Later in 1990s, several new analogous compounds $R$BiS$_2$O ($R$ = La, Pr, Nd and Yb) were synthesized, and were found to exhibit semiconducting behavior.\cite{R1121} Until recently, superconductivity was observed by electron doping via F-for-O substitution (the first rational strategy mentioned above).\cite{mizuguchi-2} Band structure calculations indicate that the prototype parent compound LaBiS$_2$O is a band insulator with an energy gap of $\sim$ 0.8 eV\cite{wxg,li,yildirim,wgt}. The conduction bands mainly consist of Bi- 6$p_x$ and 6$p_y$ orbitals which are probably responsible for the superconductivity (note that hole doping fails to induce superconductivity in the 1121 system\cite{yazici-2}). This leads to an effective two-orbitals ($p_x$ and $p_y$) model,\cite{kuroki} on which possible unconventional superconductivity was proposed.\cite{kuroki,wqh,dagotto}

We recently synthesized a novel member in the 1121 family, EuBiS$_2$F (Eu-1121), which shows a possible charge-density wave (CDW) and superconductivity without extrinsic chemical doping.\cite{zhai} As a comparison, its sister compound SrBiS$_{2}$F is a band insulator,\cite{lei,lyk} and superconductivity appears only by electron doping through a heterovalent substitution.\cite{lyk} We revealed that this Eu-1121 material was actually self doped due to Eu valence fluctuations.\cite{zhai} Albeit of relatively low doping level and presence of CDW instability, superconductivity emerges at 0.3 K. These results demonstrate that the Eu-containing bismuth sulfides bear new interesting phenomena that are distinct from other BiS$_2$-based family members. Therefore, it is worthwhile to explore its analogous compounds by the second strategy above -- structural design. In this article, we report our successful trial in synthesizing a new Eu-containing bismuth sulfide, Eu$_3$Bi$_2$S$_{4}$F$_4$ (Eu-3244), designed by incorporating an additional EuF$_2$ layer into the Eu-1121 structure (see right side of Figure ~\ref{structure}). The crystal structure is of a new type, to our knowledge, which contains thicker spacer layers of the Eu$_3$F$_4$ block in addition to the essential 2D BiS$_2$ bilayers. Unlike Eu-1121, there are two crystallographically distinct europium sites in Eu-3244. The Eu ions in the middle of the Eu$_3$F$_4$ layers were revealed to be in a mixed valence (MV) state with a valence of $v_{\text{Eu}} \sim$ +2.5 $-$ 2.75 (depending on the different measurement techniques used), while other Eu ions on the top and bottom sides of the Eu$_3$F$_4$ block layers are essentially divalent. The material shows superconductivity at 1.5 K, five times of the $T_\text{c}$ value of Eu-1121, with no CDW anomaly below 300 K. The appearance of superconductivity is explained by self electron doping into the conduction bands, owing to the Eu MV states. Implications of the results are discussed.

\section{Experimental Section}

The designed compound Eu$_3$Bi$_2$S$_{4}$F$_4$ was synthesized by a solid-state reaction method. First, EuS was presynthesized by reacting stoichiometric Eu pieces (99.9\% Alfa Aesar) and S powders (99.9995\% Alfa Aesar) at 1073 K for 10 h. Second, the stoichiometric mixtures of EuS, EuF$_2$ (99.9\% Alfa Aesar) and Bi$_2$S$_3$ (99.999\% Alfa Aesar) were sintered at 1123 K for 35 h in a sealed evacuated quartz ampoule. The sample purity were able to be improved by repeating the sintering procedure. All the operations concerning with the intermediate materials were conducted in an argon-filled glove box to avoid the contamination of water and oxygen as far as possible. Nevertheless, the final product, being black in color, is stable in air.

Powder X-ray diffraction (XRD) was carried out at room temperature on a PANalytical X-ray diffractometer (Model EMPYREAN) with a monochromatic Cu K$_{\alpha1}$ radiation. The lattice parameters were determined by a least-squares fit using Si as internal standard reference material. The crystal structure was determined by a Rietveld refinement, adopting the structural model designed and, using the program RIETAN-2000.\cite{rietan} The resultant weighted reliable factor $R_{\text{wp}}$ is 6.46\% and the goodness-of-fit parameter $S$ is 1.16, indicating validity and correctness of the refinement. High-resolution transmission electron microscope (HRTEM) images were taken at room temperature with a FEI Tecnai G$^2$ F20 scanning transmission electron microscope. The chemical composition of the new compound was verified by energy dispersive x-ray (EDX) spectroscopy affiliated to a field emission scanning electron microscope (FEI Model SIRION). The electron beam was focused on a crystalline grain (see Figure S1), and ten EDX spectra from different grains were collected.

$^{151}$Eu M\"{o}ssbauer spectra were measured for the Eu-3244 specimen by using a conventional constant acceleration drive and $\sim$ 50 mCi $^{151}$Sm$_2$O$_3$ source. The experimental spectra were analyzed by two Lorentzian lines from which values for the isomer shift ($\delta$), the spectral area of the resonance absorption lines, and the quadrupole interactions were derived. The analysis considered also the exact shape of the source emission line\cite{felner}. The velocity calibration was performed with an $\alpha$-iron foil at room temperature and the reported $\delta$ values are relative to Eu$_2$O$_3$ at room temperature.

Magnetic measurements were performed on a Quantum Design Magnetic Property Measurement System (MPMS-5) using 6.06 mg Eu-3244 sample. The temperature dependence of dc magnetic susceptibility was measured from 1.9 to 400 K. The dc magnetization was also measured as a function of magnetic field. The temperature-dependent resistivity and specific heat capacity were measured down to 0.5 K using a standard four-probe method and a relaxation technique, respectively, on a Quantum Design Physical Property Measurement System (PPMS-9) with a $^3$He refrigerator inserted. A rectangular bar (2.5$\times$1.5$\times$1.1 mm$^3$), cut from the as-prepared Eu-3244 pellet, was employed for the measurement. Gold wires ($\phi$=30 $\mu$m) were attached onto the sample's surface with silver paint, and the contact pads caused an uncertainty in the absolute values of resistivity of $\pm$20\%. The applied current was 0.5 mA. In the heat capacity measurement, the sample was cut and polished into a thin square plate (2.8$\times$2.8$\times$0.2 mm$^3$, 10.7 mg), and it was carefully placed onto the sample holder coated with a thin layer of Apiezon N-grease. The heat capacity from the sample holder and grease was deducted. All the specimens measured were taken from the same batch.

\section{Results and Discussion}

\subsection{Crystal Structure}

\begin{figure}
\includegraphics[width=10cm]{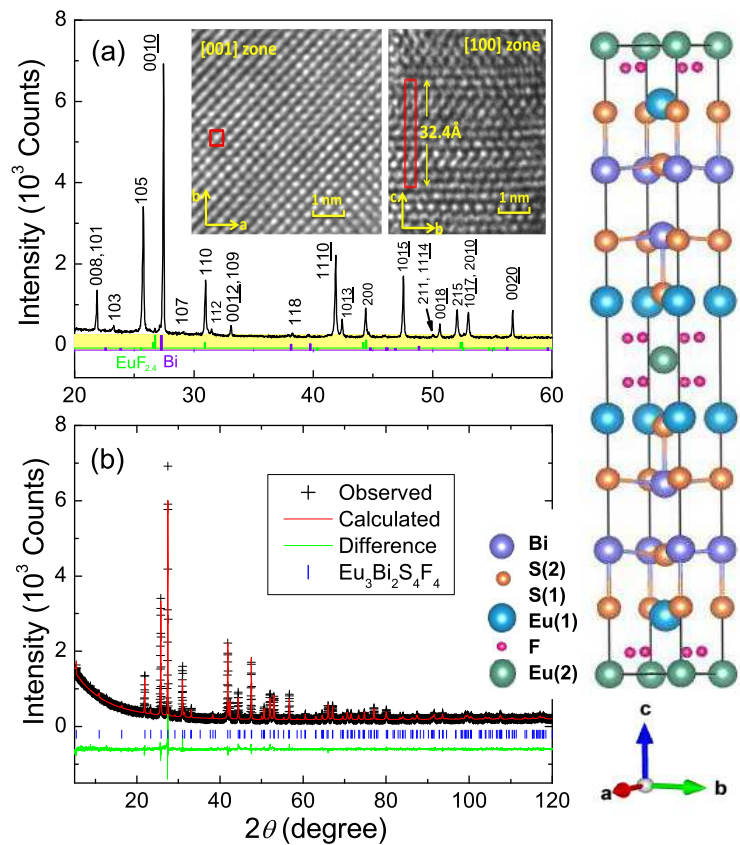}
\caption{Structural characterizations of Eu$_3$Bi$_2$S$_{4}$F$_4$ at room temperature by powder X-ray diffraction (XRD) and high resolution transmission electron microscopy (HRTEM). (a) The XRD patterns in the range of 20$^\circ \leq 2\theta \leq 60^\circ$, indexed with a body centered tetragonal unit cell whose dimensions are confirmed by the HRTEM images shown in the inset. The diffraction lines of EuF$_{2.4}$ and Bi, as labelled at the bottom, are taken from ICDD PDFs \#26-0626 and \#85-1331. (b) The Rietveld refinement profile by adopting the structure model displayed on the right side.  \label{structure}}
\end{figure}

Figure ~\ref{structure} shows structural characterizations for the Eu-3244 sample. The powder XRD patterns can be well indexed by a tetragonal cell with  $a$ = 4.077 {\AA} and $c$ = 32.433 {\AA}, except for several tiny reflections (the intensity of the strongest impurity peak is less than 3\% of that of the main phase) from Bi and EuF$_{2.4}$ impurities. The diffraction indexes ($h k l$) obey $h + k + l =$ even numbers, indicating a body-centered lattice. The unit cell is confirmed by HRTEM images of the [001] and [100] zones (see the inset of Figure ~\ref{structure}a). Compared with the cell parameters of Eu-1121 [$a$ = 4.0508(1) {\AA} and $c$ = 13.5338(3) {\AA}],\cite{zhai} the $a$ axis is merely 0.026 {\AA} larger, and one half of the $c$ axis (because of the body centered unit cell) is 2.683 {\AA} longer. The difference ($\frac{1}{2}$$c_{3244}-$$c_{1121}$) basically agrees with the thickness of a single EuF$_2$ crystallographic layer (e.g., 2.92 {\AA} in cubic EuF$_{2}$, according to ICDD PDF \#12$-$0391). The HRTEM images confirm the structure as designed, by showing the heavy elements (like Eu and Bi) at their corresponding positions. Our EDX spectroscopy experiment (see Figure S1 and Table S1) gives a chemical formula of Eu$_{3.1(4)}$Bi$_{1.8(1)}$S$_{4.4(3)}$F$_{3.7(7)}$, which is close to the stoichiometric Eu$_3$Bi$_2$S$_{4}$F$_4$. Furthermore, based on the structure model, the Rietveld refinement was very successful (note that preferential orientation of crystalline grains was included because of the plate-like morphology as seen in the inset of Figure S1), as shown in Figure ~\ref{structure}b. We thus conclude that the target new material Eu$_3$Bi$_2$S$_{4}$F$_4$, with additional EuF$_2$ layers to the Eu-1121 structure, was successfully synthesized.

The refined structural parameters of Eu-3244 are listed in Table ~\ref{tbl:structure}. Notably, there are two different crystallographic sites, 4$e$ and 2$a$ for Eu(1) and Eu(2), respectively. The Eu(1) cations are coordinated by four F$^{-}$ and five S$^{2-}$ anions, the same as all the Eu ions in Eu-1121. For the Eu(2) cations, however, the coordination atoms are eight nearest F$^{-}$ anions [there are two S(1) anions at a distance of 3.80 {\AA} from Eu(2), which are usually not considered as the ligands]. As is known, the valence of a certain cation can be evaluated by the bond valence sum (BVS).\cite{BVS0} Based on the bond distances in Table ~\ref{tbl:structure}, the Eu-BVS was calculated by the formula $\sum\text{exp}(\frac{R_{0}-d_{ij}}{0.37})$, where $R_{0}$ is empirical parameters and $d_{ij}$ represents the bond distances between Eu and its coordination anions. Since the $R_{0}$ value is 2.04 {\AA} and 1.961 {\AA}, respectively, for Eu$^{2+}$ -- F$^{-}$ and Eu$^{3+}$ -- F$^{-}$ bonds,\cite{BVS} we performed a self-consistent iteration process for the calculation. The resultant Eu-BVS values for Eu(1) and Eu(2) are +1.96(3) and +2.64(5), respectively [note that inclusion of the two S(1) will not bring an appreciable change (within 0.01) in Eu(2)-BVS]. This result indicates an unusual scenario for the Eu valence states: the Eu(1) ions are essentially divalent, but simultaneously, the Eu(2) ions are in a MV state.

\begin{table}
  \caption{Crystallographic Data of Eu$_3$Bi$_2$S$_{4}$F$_4$ from the Rietveld Refinement for the Powder X-ray Diffractions at 300 K.}
  \label{tbl:structure}
  \begin{tabular}{lllll}
      \hline \hline
    chemical formula &  & & &Eu$_3$Bi$_2$S$_{4}$F$_4$\\
    space group &  & &  &I4/$mmm$ (No. 139) \\
    $a$ (\r{A}) &  & &  &4.0771(1) \\
    $c$ (\r{A}) &  & &  &32.4330(6) \\
    $V$ (\r{A}$^{3}$) &  & &  &539.13(2) \\
    $Z$ & & &  &2 \\
    $\rho$ (g/cm$^{3}$) &  & &  &6.6413(2) \\
    $R_{\text{wp}}$ (\%) & & &  &6.46 \\
    $S$ &  & &  &1.16 \\

    \hline
    atom    & site     & $x$    & $y$      & $z$  \\
    \hline
    Eu(1)    & 4$e$     &0     & 0     & 0.5928(1)  \\
    Eu(2)    &  2$a$    &0     & 0     & 0  \\
    Bi     & 4$e$       &0     & 0     & 0.1978(1)  \\
    S(1)     & 4$e$     &0     & 0     & 0.1173(3)  \\
    S(2)     & 4$e$     &0     & 0     & 0.6988(5)  \\
    F       & 8$g$      &0     & 0.5   & 0.0392(3)  \\
    \hline
    bond distances & & &  & multiplicity \\
    \hline
    Eu(1)$-$F (\r{A}) &     &    &    2.678(5) & $\times$4   \\
    Eu(2)$-$F (\r{A}) &     &    &    2.403(6)  & $\times$8   \\
    Eu(1)$-$S(1) (\r{A})&   &     &   3.438(14)  & $\times$1   \\
    Eu(1)$-$S(2) (\r{A})&    &    &    2.990(4)   & $\times$4   \\
    \hline
    \hline
  \end{tabular}
\end{table}

\subsection{M\"{o}ssbauer Spectra}

\begin{figure}
\includegraphics[width=7.5cm]{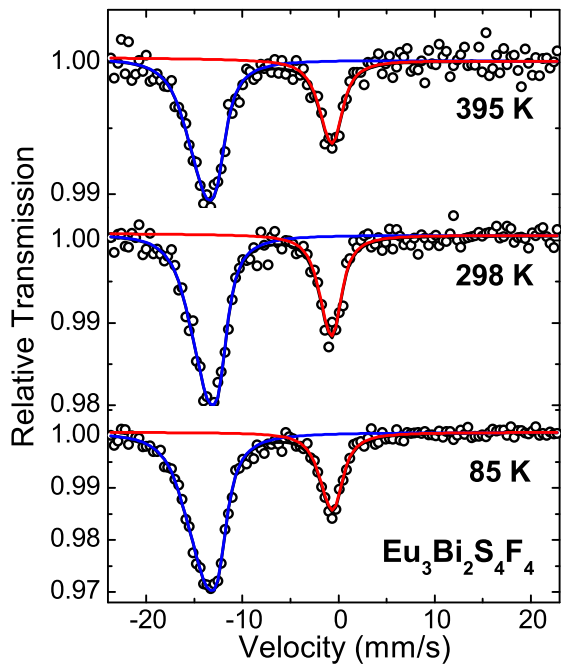}
  \caption{$^{151}$Eu M\"{o}ssbauer spectra of Eu$_3$Bi$_2$S$_{4}$F$_4$ at 85, 298 and 395 K. The blue and red lines represent the fitted curves contributed from Eu$^{2+}$ and Eu$^{3+}$ ions, respectively.
  \label{moss}}
\end{figure}

To confirm the Eu valence state in Eu-3244, the $^{151}$Eu M\"{o}ssbauer spectra were measured, as shown in Figure ~\ref{moss}. Two separate absorption lines are displayed with the isomer shifts at $\delta_1$ = $-$13.9(1) and $\delta_2$ = $-$0.8(1) mm/s, which are obviously ascribed to the resonance absorptions of Eu$^{2+}$ and Eu$^{3+}$ nuclei, respectively. With increasing the temperature up to 395 K, neither the $\delta_1$ and $\delta_2$ values, nor the absorption intensity of Eu$^{3+}$ relative to that of Eu$^{2+}$, and nor the line widths change significantly. This is in sharp contrast with the observations in Eu-1121 material where all the M\"{o}ssbauer parameters vary with temperature, in relation with Eu valence fluctuations.\cite{zhai} Since the probing time of M\"{o}ssbauer measurement is about 10$^{-9}$ s, which is much longer than the usual time scale of electron dynamics in solids, the temperature-independent isomer shifts and the line widths indicate static or quasi-static MV of Eu. Note that the isomer shifts of Eu$^{2+}$ and Eu$^{3+}$ lines are about 0.3 mm/s smaller than those of Eu-1121 at low temperatures. This may reflect difference in the hybridization between Eu-4$f$ electrons and the conduction electrons.

Table ~\ref{ms} lists the $^{151}$Eu M\"{o}ssbauer parameters at different temperatures obtained by the data fitting with consideration of the exact shape of the emission spectrum of $^{151}$Sm$_2$O$_3$ and the quadrupole interactions. Consistent with the above qualitative analysis, there is no obvious variations in isomer shifts, line widths, and also quadrupole interactions for the Eu$^{2+}$ lines, with changing temperature. Notably, no electrical quadrupole interaction is derived for the Eu$^{3+}$ absorption lines. This means that the Eu$^{3+}$ ions are indeed located at the Eu(2) site where the electric field gradient is virtually zero due to its highly symmetric coordinations. The population fraction of Eu$^{3+}$, scaled by the relative absorption areas, shows minor temperature dependence, from 0.25 (at 85 K) to 0.30 (at 395 K), which can be interpreted by a small difference in the Debye-Waller factors between Eu$^{3+}$ and Eu$^{2+}$. Similar observation was reported in Eu$_2$CuS$_3$, in which the Debye temperature of Eu$^{3+}$ is higher than that of Eu$^{2+}$.\cite{Eu2CuS3} In this circumstance, low-temperature data should more accurately reflect the population fraction. Taken the 85-K result, i.e., Eu$^{3+}$/(Eu$^{2+}$+Eu$^{3+}$) = 0.25(1), and with all the Eu$^{3+}$ ions occupied Eu(2) site [simultaneously, the Eu(1) ions are purely divalent], the Eu(2) valence, $v_{\text{Eu2}}$, can be easily estimated to be +2.75(3), which is not far from the Eu(2)-BVS value of +2.64(5). One may expect that the 'real' $v_{\text{Eu2}}$ value could be somewhat lower, if considering a little higher recoil-free fraction for Eu$^{3+}$ ions, and if the M\"{o}ssbauer data were collected at lower temperatures. Therefore, the unusual Eu valence states obtained by BVS approach are confirmed by M\"{o}ssbauer techniques.

\begin{table}
  \caption{$^{151}$Eu M\"{o}ssbauer parameters\textsuperscript{\emph{a}} of Eu$_3$Bi$_2$S$_{4}$F$_4$ by data fitting.}
  \label{ms}
  \begin{tabular}{llll}
  \hline \hline
    Temperature (K)    & 85       & 298     & 395  \\
   \hline
    $\delta_1$ (mm/s)    &  $-$13.87(3)  &$-$13.62(6)  & $-$13.87(7) \\
    $\delta_2$ (mm/s)    &  $-$0.78(5)   &$-$0.85(6)    & $-$0.82(9) \\
    $\Gamma$ (mm/s)      & 2.3(1)    &2.1(1)       & 2.3(2)  \\
    $\frac{1}{4}e^{2}qQ$ (Eu$^{2+}$)  & 3.1(1)      &2.7(2)     & 2.7(2) \\
    $\frac{1}{4}e^{2}qQ$ (Eu$^{3+}$)  & --      &--     & -- \\
    Eu$^{3+}$ fraction      & 0.25(1)   &0.29(1)    & 0.30(1)  \\
    \hline
    \hline
  \end{tabular}

  \textsuperscript{\emph{a}} $\delta_1$ and $\delta_2$ denote the isomer shift for Eu$^{2+}$ and Eu$^{3+}$, respectively. $\Gamma$ refers to the full linewidth at half maximum. $\frac{1}{4}e^{2}qQ$ represents the quadrupole interactions. The population fraction of Eu$^{3+}$ was scaled by the relative absorption areas.
\end{table}

\subsection{Magnetic Susceptibility}

\begin{figure}
  \includegraphics[width=10cm]{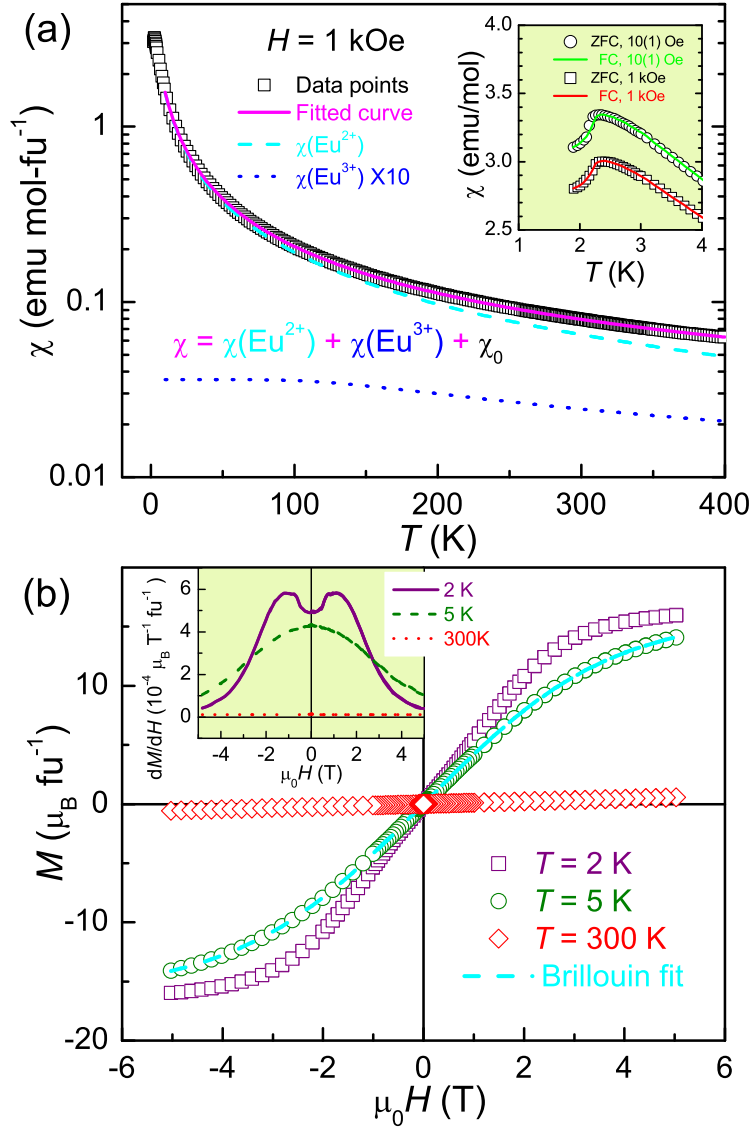}
  \caption{Magnetic measurements on Eu$_3$Bi$_2$S$_{4}$F$_4$. (a) Temperature dependence of dc magnetic susceptibility. Note that a logarithmic scale is applied for showing the contribution from Eu$^{3+}$ clearly. The inset zooms in the low-temperature data in which an antiferromagnetic transition at 2.3 K can be seen. The small inconsistency in the value of $\chi$ comes from the deviation of magnetic field from 10 Oe. (b) Field dependence of magnetization at 2, 5, and 300 K. The dotted line on the 5-K curve is a Brillouin fit. The inset plots the derivative of magnetization.}
  \label{mag}
\end{figure}

Figure ~\ref{mag}a shows the temperature dependence of magnetic susceptibility (in emu/mol-fu, where fu refers to formula unit), $\chi(T)$, for the Eu-3244 sample. The $\chi(T)$ data well follow the extended Curie-Weiss law, $\chi=\chi_{0}+C/(T+\theta_{\text{N}})$, where $\chi_{0}$ denotes the temperature independent components of susceptibility, $C$ the Curie constant, and $\theta_{\text{N}}$ the paramagnetic Neel temperature. The Curie-Weiss fitting gives an average effective moment of 7.28 $\mu_{\text{B}}$ Eu-atom$^{-1}$, which is substantially less than the expected value for Eu$^{2+}$ ions (7.94 $\mu_{\text{B}}$ Eu-atom$^{-1}$). This indicates existence of Eu$^{3+}$ ions which show Van Vleck paramagnetism due to spin-orbit interaction.\cite{vv1,vv2} The Van Vleck magnetic susceptibility is a function of the spin-orbit coupling constant $\lambda$ as well as temperature $T$, i.e., $\chi_{\text{VV}}\propto f(\lambda,T)$.\cite{vv1,vv2,nagata} With $\tilde{v}_{\text{Eu}}$ denotes the average Eu valence [consequently, the Eu$^{2+}$ content is $(9-3\tilde{v}_{\text{Eu}})$ per fu, and the Eu$^{3+}$ content is $(3\tilde{v}_{\text{Eu}}-6)$ per fu], therefore, a more accurate fitting of the $\chi(T)$ data in a wide range of 10 K $\leq T \leq$ 400 K was performed by equation,
\begin{equation}
\chi(T)=\chi(\mathrm{Eu^{2+}})+\chi(\mathrm{Eu^{3+}})+\chi_{0}, \label{eqn:example}
\end{equation}
where $\chi(\mathrm{Eu^{2+}})=C_{2+}(9-3\tilde{v}_{\text{Eu}})/(T+\theta_{\text{N}})$ and $\chi(\mathrm{Eu^{3+}})=C_{3+}(3\tilde{v}_{\text{Eu}}-6)f(\lambda,T)$. With the unit of emu mol-fu$^{-1}$ for $\chi(T)$, $C_{2+}$ and $C_{3+}$ are settled to be 7.875 and 0.125 emu mol-fu$^{-1}$ K$^{-1}$, respectively, and there are actually only four parameters ($\tilde{v}_{\text{Eu}}, \theta_{\text{N}}, \lambda$ and $\chi_{0}$) to be fitted. The fitting was very successful, which gives rise to the following results. (1) $\tilde{v}_{\text{Eu}}$ = 2.167(1), meaning 2.50 Eu$^{2+}$ and 0.50 Eu$^{3+}$ in a unit formula of Eu-3244. This conclusion basically agrees with the Eu-BVS and M\"{o}ssbauer results above. (2) The spin-orbit coupling constant $\lambda$ is fitted to be 419$\pm$19 K, which is a reasonable value for the related europium compounds.\cite{nagata}  (3) The fitted $\chi_{0}$ value is as high as 0.0120(1) emu mol-fu$^{-1}$, suggesting an exceptionally large Pauli paramagnetic contribution. This means very high density of state at the Fermi level ($E_{\text{F}}$), like the case in Eu-1121 which shows substantial hybridizations between Eu-4$f$ and conduction bands.\cite{zhai} The heat capacity result below also supports this speculation. (4) The paramagnetic Neel temperature $\theta_{\text{N}}$ is +2.63 K, suggesting a dominant antiferromagnetic interactions among the Eu$^{2+}$ spins. This is consistent with the antiferromagnetic-like transition at 2.3 K, as shown in the inset of Figure ~\ref{mag}a.

The above analysis on $\chi(T)$ is actually based on the presumption that $\tilde{v}_{\text{Eu}}$ does not change with temperature. So, it is necessary to check the $\tilde{v}_{\text{Eu}}$ value at a certain temperature, preferably at sufficiently low temperatures which reflects ground-state properties. The field-dependent magnetization, $M(H)$, shown in Figure ~\ref{mag}b, just gives the information. At 300 K, the magnetization increases slowly and linearly with the applied field, in accordance with the small susceptibility. At low temperatures, however, the $M(H)$ curves deviate from linearity, and tend to saturate at high magnetic fields. The $M(H)$ curve at 5 K can be well fitted by a Brillouin function with consideration of Weiss molecular field $B_{\text{mf}}=\xi M$. The fitted saturation magnetization is 16.78 $\mu_{\text{B}}$ fu$^{-1}$, corresponding to 2.40 Eu$^{2+}$ and 0.60 Eu$^{3+}$ in a unit formula of Eu-3244. The result is in principle consistent with the $\chi(T)$ fitting result, and also meets the conclusion from the BVS and M\"{o}ssbauer-spectrum evaluations. Besides, the fitted parameter $\xi$ is negative ($-$0.09), conforming to the positive value of
$\theta_{\text{N}}$ and the antiferromagnetic transition at 2.3 K. The $M(H)$ curve at 2 K cannot be well fitted by a Brillouin function because it is in a magnetically ordered state. The derivative of magnetization shown in the inset of Figure ~\ref{mag}b indiactes a metamagnetic-like transition occurs around the field $\mu_{0}H$ = 1 T below the Neel temperature.

\subsection{Electrical Resistivity}

\begin{figure}
\includegraphics[width=10cm]{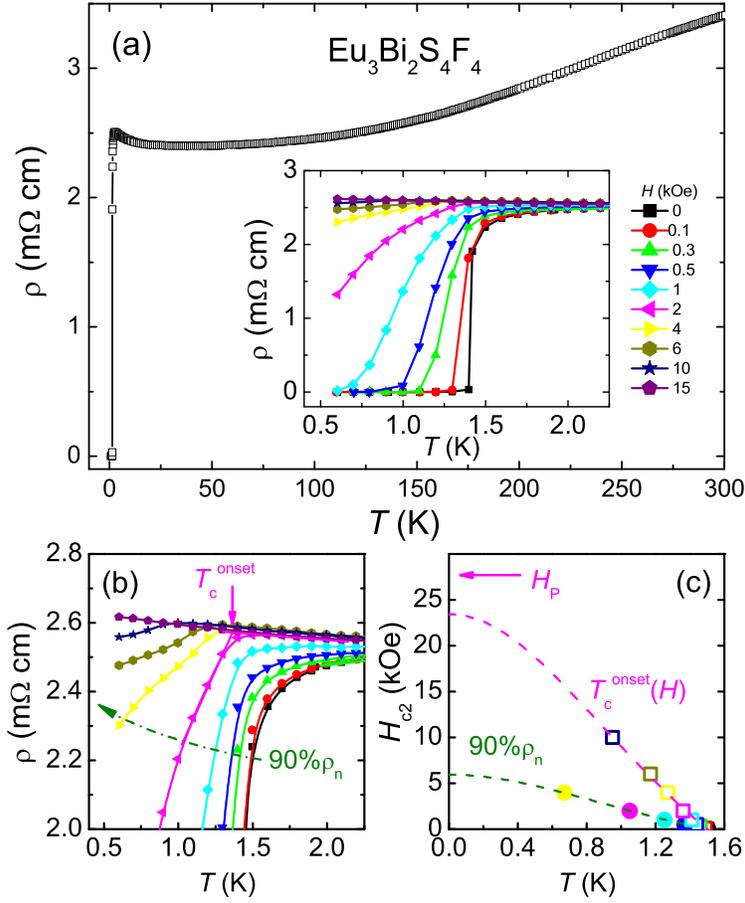}
\caption{Superconductivity in Eu$_3$Bi$_2$S$_{4}$F$_4$. (a) The temperature dependence of resistivity. The inset shows the superconducting transition under various magnetic fields. (b) An enlarged plot of the temperature dependence of magnetoresistivity, from which the upper critical fields, $H_{\text{c2}}$, can be derived with two criteria. (c) The superconducting $H_{\text{c2}}-T$ phase diagram. The dashed lines are fitted curves based on Ginzburg-Landau theory. The Pauli limiting field, $H_\text{P}$ is marked. \label{res}}
\end{figure}

Figure ~\ref{res}a shows the temperature dependence of resistivity, $\rho(T)$, for the Eu-3244 polycrystalline sample. It exhibits metallic conduction with positive temperature coefficient of resistivity (TCR) above 35 K, albeit without extrinsic doping. This is obviously due to the self doping effect arising from the Eu MV, like the case in Eu-1121. The positive TCR is in contrast with the negative TCR in the electron-doped Sr$_{1-x}$La$_x$BiS$_{2}$F system even for $x>$ 0.5.\cite{lyk,sakai,lyk2} The latter was interpreted by Anderson localization,\cite{sakai} because the extrinsic La/Sr substitution brings about disorder potentials in the conducting BiS$_2$ bilayers. In Eu-3244 and Eu-1121 systems, no extrinsic chemical doping is applied, and the Anderson localization effect is largely reduced. Compared with Eu-1121, however, it is remarkable that Eu-3244 does not show any CDW-like anomaly below 300 K.

Below 1.5 K, the resistivity drops abruptly, and it achieves zero at 1.4 K (see the inset), indicating a superconducting transition. The narrow superconducting transition width ($<$ 0.1 K) at zero field suggests bulk superconductivity (note that the measured sample is nearly monophasic polycrystal). The $T_\text{c}$ value is nearly five times of that in Eu-1121, and is comparable to that in Sr$_{1-x}$La$_x$BiS$_{2}$F system with $x$ = 0.4.\cite{lyk2}

The superconductivity was confirmed by the magnetoresistance measurement.\bibnote{We were not able to test the Meissner effect by dc magnetic susceptibility measurement due to the limit of the lowest temperature achieved. However, our ac susceptibility measurement down to lower temperatures shows a sharp decrease below 1.4 K, confirming the superconductivity. Details will be reported elsewhere.}  With applying magnetic fields, the superconducting transition shifts down to lower temperatures. Moreover, the transition width is broadened, which is the characteristic of strongly anisotropic superconductors in which magnetic flux flow produces extra resistance. The upper critical field, $H_{\text{c2}}$, can be obtained from the $\rho(T,H)$ data. If employing the conventional criterion of 90\% of the normal-state resistivity to determine $T_{\text{c}}(H)$, $H_{\text{c2}}$ could be underestimated because of the severe broadening of the superconducting transition. Hence, we made an additional criterion to determine $H_{\text{c2}}$ by defining $T_{\text{c}}^{\text{onset}}(H)$ (see Figure ~\ref{res}b). The $H_{\text{c2}}(T)$ data satisfy the Ginzburg-Landau law,\cite{GL} $H_{\text{c2}}(T) = H_{\text{c2}}(0)(1-t^{2})/(1+t^{2})$, where $t$ denotes a reduced temperature of $T/T_{\text{c}}$. The upper critical field at zero temperature, $H_{\text{c2}}(0)$, is then estimated to be 5.9 kOe and 23.4 kOe, respectively, for the different criteria employed. The latter $H_{\text{c2}}(0)$ value is close to $H_{\text{P}}$ = 18.4 $T_{\text{c}}\approx$ 27.6 kOe, suggesting Pauli limited behavior in this system.

\subsection{Specific Heat}

\begin{figure}
\includegraphics[width=10cm]{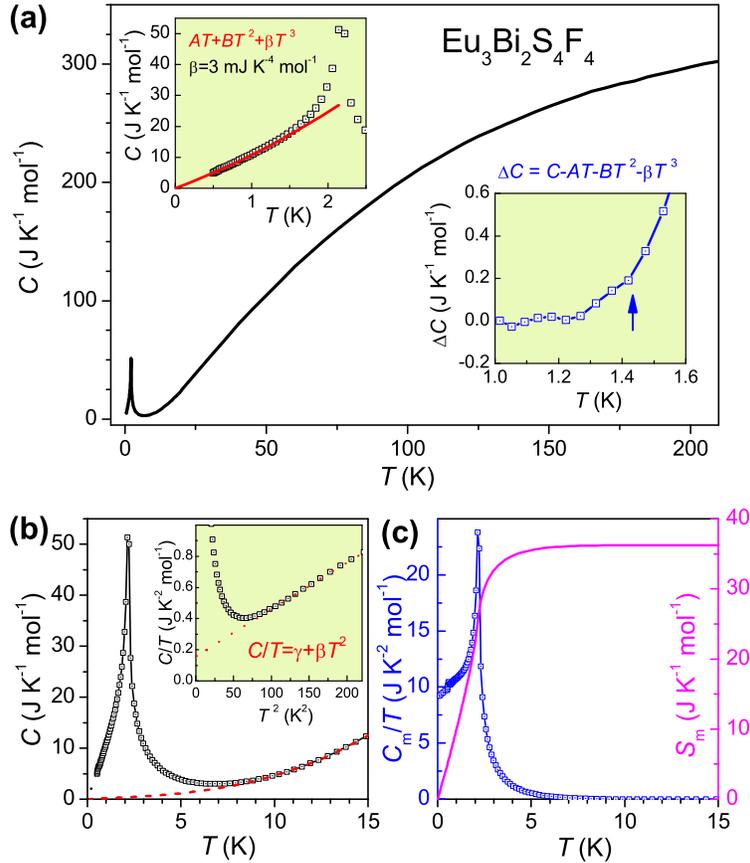}
\caption{\label{fig5} Specific heat capacity for Eu$_3$Bi$_2$S$_{4}$F$_4$. (a) Temperature dependence of specific heat capacity, $C(T)$, from 0.5 to 210 K. The upper left inset zooms in the low-temperature $C(T)$ data. The solid line is a fit by $C = A T + B T^2+\beta T^3$, where $\beta$ was fixed to be 3.00 mJ K$^{-4}$ mol$^{-1}$. The lower right inset shows $\Delta C$, defined by $C-AT-BT^2-\beta T^3$, as a function of temperature. (b) An enlarged plot of $C(T)$ from 0.5 to 15 K. The dashed line represents the sum of electronic and lattice contributions. The inset plots $C/T$ vs. $T^2$, where the dashed line obeys $C/T = \gamma + \beta T^2$. (c) $C_{\text{m}}/T$ (right axis, where $C_{\text{m}}$ is the difference between the two sets of data in (b), representing the magnetic contribution to the specific heat) and magnetic entropy $S_\text{m}$ (left axis) as functions of temperature.\label{SH}}
\end{figure}

Figure~\ref{SH}a shows the temperature dependence of specific
heat, $C(T)$, for the Eu-3244 sample. The $C(T)$ data at high temperatures tend to saturate above 300 J K$^{-1}$ mol$^{-1}$, close to the
Dulong-Petit value of 3$NR$ = 39$R$ = 324 J K$^{-1}$ mol$^{-1}$, where $N$ counts the number of elements per fu, $R$ is the gas constant. A sharp $\lambda$-shape peak at 2.2 K is observed (see the magnified plot in the upper left inset), coinciding with magnetic measurement which indicates an antiferromagnetic ordering of the Eu$^{2+}$ spins at 2.3 K. The divergence in $C(T)$ at 2.3 K confirms the long-range magnetic ordering, which is in contrast with the hump-shape peak for Eu-1121 where a spin glass state is likely to formed.\cite{zhai} The low-temperature $C(T)$ data ($T <$ 2 K) are predominately from magnetic contributions, which can be fitted by the sum of a linear term and a quadratic term. The quadratic term can be ascribed to the excitations of 2D antiferromagnetic spin waves.\cite{joshua} The linear term, which generally appears in a spin-glass state,\cite{tari} implies existence of spin glass that probably occurs in the Eu(2) sublattice due to the Eu mixed valence. The large magnetic signal of $\sim$ 20 J K$^{-1}$ mol$^{-1}$ prevents the direct observation of a possible jump in $C(T)$, estimated to be $\sim$ 0.2 J K$^{-1}$ mol$^{-1}$, associated with the superconducting transition around 1.5 K. Nevertheless, we indeed observed a kink at 1.43 K (see the lower right inset of Figure 5a) that could be related to the superconducting transition, when subtracting the contributions from magnetism (the linear and quadratic terms), electrons and lattice.

Similar to the $C(T)$ analysis for Eu-1121,\cite{zhai} the different contributions from conduction electrons ($C_{\text{el}}$), crystalline lattice ($C_{\text{lat}}$) and Eu magnetism
($C_{\text{m}}$) can be separated, assuming that $C_{\text{m}}$ is negligibly small above 10 K. The $C/T$ vs $T^2$ plot, shown in the inset of Figure~\ref{SH}b gives $\gamma$ = 159 mJ K$^{-2}$ mol$^{-1}$ and $\beta$ = 3.00 mJ K$^{-4}$ mol$^{-1}$. The Debye temperature is then obtained by $\theta_{\text{D}} = [(12/5)NR\pi^{4}/\beta]^{1/3}$ = 203 K, which is almost identical to that of Eu-1121.\cite{zhai} The fitted $\gamma$ value is also close to that of Eu-1121 based on per mol-Bi (because of Bi-6$p$ electrons mainly contribute the conduction bands). Using the formula $N(E_{\text{F}}) = 3\gamma /(\pi k_{\text{B}})^{2}$, the density of state at $E_{\text{F}}$ estimated is as high as 65 eV$^{-1}$ fu$^{-1}$, which explains the unusually large Pauli magnetic susceptibility above.

The magnetic contribution $C_{\text{m}}$ was then obtained by removing the sum of $C_{\text{lat}}$ (= $\beta T^3$) and $C_{\text{el}}$ (= $\gamma T$), as shown in Figure~\ref{SH}b. Consequently, the magnetic entropy released can be calculated by the integration $S_{\text{m}}=\int_{0}^{T} (C_{\text{m}}/T)dT$ (see Figure~\ref{SH}c). We obtained $S_{\text{m}}$ = 36.2 J K$^{-1}$ mol$^{-1}$ at 15 K, equivalent to 2.09 times of $R$ln($2S+1$) ($S$=7/2 for Eu$^{2+}$). It is noted that the $S_{\text{m}}$ value could be underestimated because short-range spin ordering may extend to higher temperatures above 15 K. Remember in mind that $S_{\text{m}}$ refers to the change in magnetic entropy, the value 2.09 only counts the Eu$^{2+}$ ions that order magnetically at low temperatures. Therefore, the specific heat data do not contradict with the conclusion of Eu mixed valence above. The unexpectedly low $S_{\text{m}}$ suggests spin glass state for the Eu(2) ions, consistent with the implication of the low-temperature $C(T)$ behavior and, in conformity with the mixed valence state exclusively at the Eu(2) site.

\subsection{Discussion}

Let us first discuss the issue of Eu MV. As we know, europium is a special rare earth element that frequently shows MV due to the stability of half filling of 4$f$ orbitals for divalent Eu$^{2+}$ ions (usually trivalent rare earth is the stable state). Basically, there are two distinct categories for the Eu MV according to its dynamicity. The first MV scenario is homogeneous, in the sense that the mixed-valent Eu ions are indistinguishable in a short time (e.g., 10$^{-9}$ s). This MV behavior is characterized by valence fluctuations due to hybridization between Eu-4$f$ electrons with the conduction electrons, as exemplified by EuNi$_2$P$_2$.\cite{EuNi2P2}. Obviously, it occurs only in a material that all the Eu ions occupy the crystallographically equivalent sites. In contrast, if the Eu ions with different valences selectively occupy the different crystallographic sites, the MV will be static in nature. Such a scenario is called inhomogeneous MV, which is equivalent to static charge ordering. Examples of this category are shown in Eu$_2$CuS$_3$\cite{Eu2CuS3}, Eu$_2$SiN$_3$,\cite{Eu2SiN3} and Eu$_3$F$_4$S$_2$\cite{Eu3F4S2}, etc., where Eu$^{2+}$ and Eu$^{3+}$ occupy their individual sites. Note that, in the homogeneous Eu-MV state, $v_{\text{Eu}}$ usually increases remarkably with decreasing temperature, because of dominant thermal fluctuations (rather than quantum mixing) between the two integer-valence states.\cite{steglich} More frequently, a first-order structural transition occurs accompanying with an abrupt change in $v_{\text{Eu}}$. In such cases, the MV becomes inhomogeneous at low temperatures because of static or quasi-static character of MV, as evidenced by two separate lines in the $^{151}$Eu M\"{o}ssbauer spectra.\cite{EuCu2Si2,Eu111}

Compared with the conventional scenarios for Eu MV described above, the Eu-containing bismuth chalcogenides, including Eu-1121 and Eu-3244, exhibit unusual MV behaviors. The Eu valence in Eu-1121 was found to be nearly temperature-independent down to 2 K.\cite{zhai} This has been interpreted by Eu valence fluctuations that couple with the dynamic CDW in the BiS$_2$ bilayers. However, the situation is different here for Eu-3244. There are two crystallographically different sites for the Eu ions. Unlike the aforementioned examples that different Eu sites are occupied by Eu$^{2+}$ and Eu$^{3+}$, respectively, the Eu(2) ions at the 2$a$ site are in a MV state, while the 4$e$-site Eu(1) ions are essentially divalent. Our $^{151}$Eu M\"{o}ssbauer spectra clearly show static or quasi-static MV up to 395 K, and the Eu$^{3+}$ ions occupy the Eu(2) site. Since the Eu-4$f^{7}$ narrow band touches the Fermi level of the conduction band in Eu-1121,\cite{zhai} it is plausible that, because of stronger crystal field for Eu(2), the Eu(2)-4$f^{7}$ band is lifted further in Eu-3244. Consequently, the Eu(2) ions are in a MV state, driven by the electrons transfer from an Eu(2)-4$f$ orbital to the conduction bands. On the other hand, the Eu(1)-4$f^{7}$ band is lowered due to the relatively weak crystal field [note that the Eu(1)-BVS value is substantially smaller than the Eu-BVS in Eu-1121]. This leads to a fully filled Eu(1)-4$f^{7}$ band, i.e., the Eu(1) ions are divalent. Future band-structure calculations are expected to clarify this explanation.

Owing to the Eu(2) MV with $v_{\text{Eu2}} \approx$ 2.6(1), the conduction bands of BiS$_{2}$ bilayers should be self doped with the transferred electrons from the Eu-4$f$ states, similar to the case in Eu-1121.\cite{zhai} The doping level is estimated to be $x \sim$ 0.30(5) electrons per Bi-atom, assuming that stoichiometry is conserved. This explains the appearance of superconductivity as well as the metallic conduction in the undoped Eu-3244. Here we note that similar self doping, via an internal charge transfer from 3$d$ transition metal ions (V and Ti), leads to high-temperature superconductivity in iron-based materials.\cite{cao,sun} Thus, the direction of self doping deserves further attention.

It is unexpected that no signature of CDW transition appears in Eu-3244. In general, the enhanced 2D characteristic in Eu-3244 would further stabilize the potential CDW. Therefore, the obvious difference in Eu MV could be relevant. In Eu-3244, the Eu(2) MV occurs in the middle layers of the Eu$_3$F$_4$ block, farther from the BiS$_2$ bilayers. In addition, the Eu(2) ions show much higher valence, leading to a different pattern of charge ordering. These two factors seem to be against the possible coupling between the CDW of BiS$_2$ layers and the Eu MV.

Since CDW usually competes with superconductivity, the absence of CDW in Eu-3244 may explain the relatively high $T_\text{c}$ value (fivefold of the $T_\text{c}$ value in Eu-1121). It is also possible, of course, that the pronounced two dimensionality in Eu-3244 may enhance the superconductivity. Considering the relatively low doping level, one may expect that a higher $T_\text{c}$ could be realized by extrinsic chemical doping and/or with applying high pressures. The latter is very expectable since high pressure was found to be very effective to enhance the $T_\text{c}$ value in BiS-based materials,\cite{kotegawa,wolowiec,tomita,awana1} and high pressure may push the parent compound toward metallic conduction.\cite{awana2} Also, in this regard, exploration of new superconductors with the 3244-type structure is of potential interest.

\section{Conclusion}

We have discovered a novel europium bismuth sulfofluoride superconductor, Eu$_3$Bi$_2$S$_{4}$F$_4$, via a structural-design approach by introducing an additional EuF$_2$ layer into the existing Eu$_2$F$_2$ spacer layers in EuBiS$_{2}$F. The crystal structure, featured with the existence of two crystallographically distinct europium sites, was determined by high-resolution transmission electron microscopy as well as X-ray diffractions. We investigated the Eu valence state by: (1) structural probing with calculation of Eu-BVS; (2) internal and local probing with $^{151}$Eu M\"{o}ssbauer spectroscopy; and (3) bulk probing with analysis of magnetic properties. The results of these three probes consistently point to an anomalous Eu valence state in which the Eu(1)-site cations are essentially divalent, whilst the Eu(2)-site cations are in a mixed valence state with a valence of $\sim$ +2.6(1). The mixed valence allows charge transfer, which realizes electron doping into the conduction bands that are mainly composed of Bi- 6$p_x$ and 6$p_y$ orbitals. That is why we observed metallic conduction and superconductivity in the undoped Eu$_3$Bi$_2$S$_{4}$F$_4$. Our work could inspire further exploration of new superconductors, especially those with self doping, in BiS$_2$-based and other families.

\begin{acknowledgement}
This work was supported by NSF of China (Grants No. 11190023 and 90922002), the National Basic Research Program of China (Grant Nos. 2010CB923003 and 2011CBA00103), and the Fundamental Research Funds for the Central Universities of China (Grant No. 2013FZA3003).

\end{acknowledgement}

\begin{suppinfo}
Figure S1 (showing the energy dispersive X-ray spectrum on a crystalline grain of Eu$_3$Bi$_2$S$_{4}$F$_4$), Table S1 (listing the result of the energy dispersive X-ray sepctroscopy), and CIF file of the crystal structure data of Eu$_3$Bi$_2$S$_{4}$F$_4$.
\end{suppinfo}

\bibliography{3244_Zhai}

\begin{mcitethebibliography}{60}
\bibitem[Not()]{Note-2}
We were not able to test the Meissner effect by dc magnetic susceptibility
  measurement due to the limit of the lowest temperature achieved. However, our
  ac susceptibility measurement down to lower temperatures shows a sharp
  decrease below 1.4 K, confirming the superconductivity. Details will be
  reported elsewhere.\relax
\mciteBstWouldAddEndPunctfalse
\mciteSetBstMidEndSepPunct{\mcitedefaultmidpunct}
{}{\mcitedefaultseppunct}\relax
\EndOfBibitem
\end{mcitethebibliography}

\end{document}